\documentclass{predoc2}
\usepackage{graphicx}

\newbox{\myorcidaffilbox}
\sbox{\myorcidaffilbox}{\large\includegraphics[height=1.7ex]{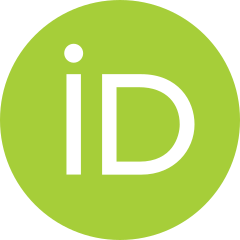}}
\newcommand{\orcidaffil}[1]{%
  \href{https://orcid.org/#1}{\usebox{\myorcidaffilbox}}}

\usepackage{csquotes}

\title{Computational modelling of biological systems now and then: revisiting tools and visions from the beginning of the century}

\author[1]{Axel Loewe\orcidaffil{0000-0002-2487-4744}\thanks{Corresponding author, publications@ibt.kit.edu}}
\author[2]{Peter J Hunter\orcidaffil{0000-0001-9665-4145}}
\author[3,4]{Peter Kohl\orcidaffil{0000-0003-0416-6270}}

\affil[1]{Institute of Biomedical Engineering, Karlsruhe Institute of Technology (KIT), Karlsruhe, Germany}
\affil[2]{Auckland Bioengineering Institute, University of Auckland, Auckland, New Zealand}
\affil[3]{Institute for Experimental Cardiovascular Medicine, University Heart Center Freiburg $\cdot$ Bad Krozingen, Freiburg, Germany}
\affil[4]{Faculty of Medicine, University of Freiburg, Freiburg, Germany}

\makeatletter
\def\runauthor{Loewe, Hunter, Kohl}
\def\runtitle{Computational modelling of biological systems now and then}
\makeatother

\begin{document}

\maketitle
\thispagestyle{empty}

\begin{abstract}Since the turn of the millennium, computational modelling of biological systems has evolved remarkably and sees matured use spanning basic and clinical research. While the topic of the peri-millennial debate about the virtues and limitations of ``reductionism \& integrationism'' seems less controversial today, a new apparent dichotomy dominates discussions: mechanistic vs. data-driven modelling. In light of this distinction, we provide an overview of recent achievements and new challenges with a focus on the cardiovascular system. Attention has  shifted from generating a universal model of \textit{the} human to either models of individual humans (digital twins) or entire cohorts of models representative of clinical populations to enable in silico clinical trials. Disease-specific parameterisation, inter-individual and intra-individual variability, uncertainty quantification as well as interoperable, standardised, and quality-controlled data are important issues today, which call for open tools, data and metadata standards, as well as strong community interactions.

The quantitative, biophysical, and highly controlled approach provided by in silico methods has become an integral part of physiological and medical research. In silico methods have the potential to accelerate future progress also in the fields of integrated multi-physics modelling, multi-scale models, virtual cohort studies, and machine learning beyond what is feasible today. In fact, mechanistic and data-driven modelling can complement each other synergistically and fuel tomorrow’s artificial intelligence applications to further our understanding of physiology and disease mechanisms, to generate new hypotheses and assess their plausibility, and thus to contribute to the evolution of preventive, diagnostic, and therapeutic approaches.
\end{abstract}

\section{Embracing the Mountain and Village Views}

The pursuit of understanding biological systems through computational modelling \& simulation (M\&S) makes use of a multitude of approaches, each providing a specific lens through which the intricate mechanisms governing life can be observed. In 2000, Kohl, Noble, Hunter \& Winslow described tools and visions for computational modelling of biological systems in the 21\textsuperscript{st} century~\cite{kohl00}. At the turn of the millennium, they likened their exploration to the contrasting perspectives of the mountain and village views, introduced in an old Chinese parable.

\begin{displayquote}
\emph{The wise man walks from the village to the top of the nearby mountain and, after a brief and peaceful rest, strides back to the village. There he stays for a short while, before he returns to the mountain, and so on. Asked why he does this, he replies that he wants to understand his people. But when he dwells inside the village, he can’t see the whole of it, and when he is on the summit, he is out of touch with the villagers. So he continues his pilgrimage for eternity.}
\end{displayquote}

Much like an observer standing atop a mountain, computational modellers often aim for a panoramic vista, seeking overarching principles and comprehensive insights that capture the concepts of biological phenomena. This integrative perspective offers a broad overview, enabling the identification of fundamental principles governing complex biological systems. However, it also entails abstraction, potentially overlooking finer nuances and intricacies inherent in the biological fabric.

Conversely, akin to an inhabitant of a village intimately familiar with its every nook and cranny, certain modelling approaches delve deeply into specific biological mechanisms, exploring their intricacies with unparalleled detail. This reductionist close-up view provides an in-depth and potentially mechanistic understanding of localised processes, allowing for precise examination and manipulation. Yet, it may come at the cost of losing sight of the interconnectedness and holistic behaviour exhibited by the broader biological system.

The discourse between these contrasting perspectives --- the panoramic vantage point from the mountain (integrationism) and the intimate familiarity within the village (reductionism) --- has settled to a certain degree and the valuable contributions of both views are appreciated now, while a new apparent dichotomy emerges: data-driven vs. mechanistic modelling. In this article, we embark on a journey that tries to navigate these contrasting views, exploring how they diverge, converge, and synergise in advancing our understanding of physiology, i.e. the ``logic of life'' with a focus on the cardiovascular system. This work covers both basic research about fundamental physiology as well as applications of in silico models to solve real-world problems as we believe they are closely intertwined: As one applies knowledge to solve problems, one learns about shortcomings of the fundamental theory, especially in complex (e.g., biological systems). 
\begin{figure}[!ht]
\centering\includegraphics[width=\columnwidth]{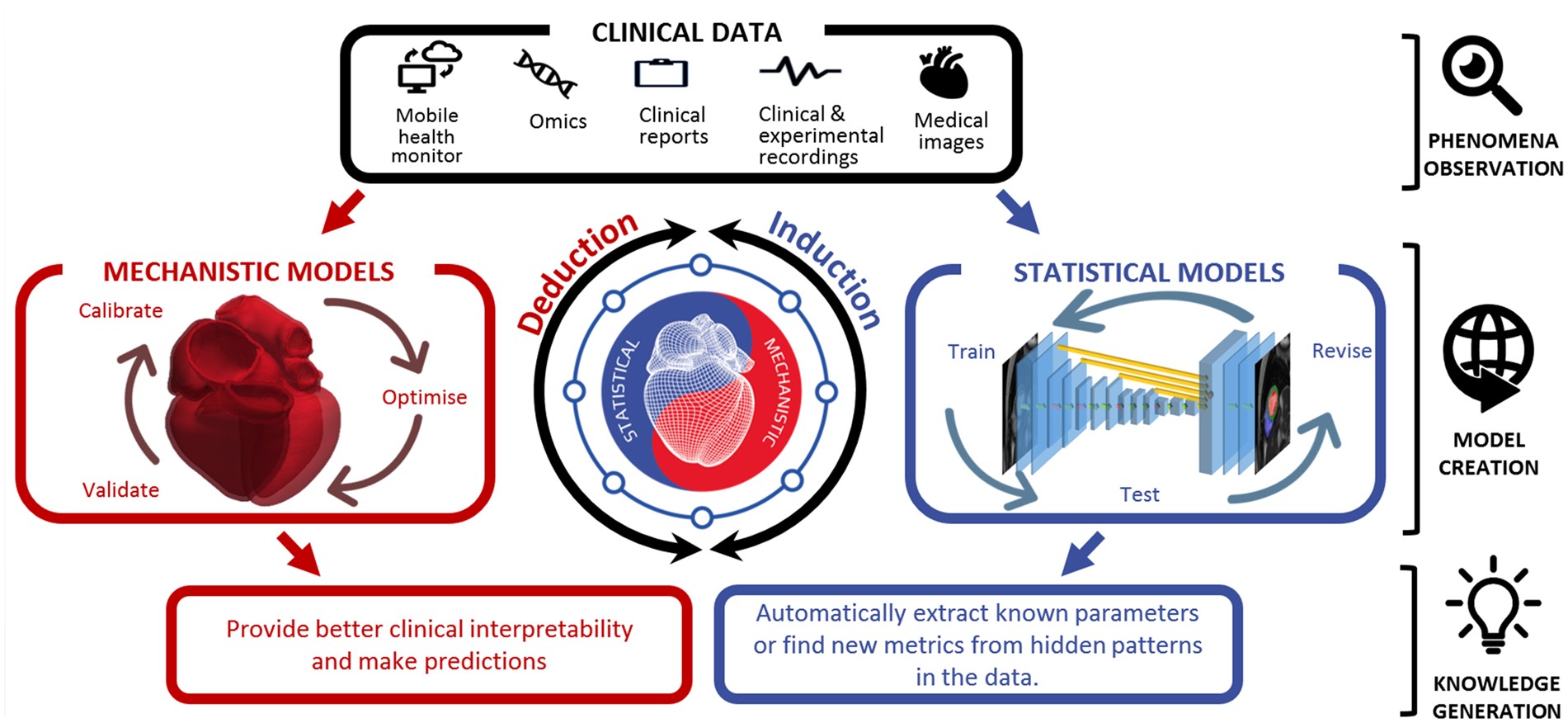}
\caption{Synergy of mechanistic and statistical (data-driven) models. Reproduced from Corral-Acero et al.~\cite{Corral-Acero-2020-ID13459} under the Creative Commons Attribution License 4.0.}
\label{fig:synergy}
\end{figure}

\section{The Millennium View from Today's Perspective}
The vision presented by Kohl, Noble, Winslow \& Hunter~\cite{kohl00} has proven to be remarkably accurate overall. Today, computational M\&S have matured significantly, witnessing widespread adoption in both basic and clinical research, as evident in the increasing number of publications utilising in silico tools. The ratio of PubMed-listed cardiology studies using M\&S is continuously rising and has multiplied five-fold since 1990 (2023: 2.2\%)\footnote{Search query: \textit{(atria*[Title/Abstract] OR ventric*[Title/Abstract] OR cardi*[Title/Abstract]) AND ('comput* model'[Title/Abstract] OR 'mathematical model'[Title/Abstract] OR 'in silico'[Title/Abstract])} vs. \textit{(atria*[Title/Abstract] OR ventric*[Title/Abstract] OR cardi*[Title/Abstract])}}. The advantages initially highlighted, including faster and more cost-effective research, arbitrary resolution within the model's scope, and enhanced availability, remain true and have advanced over time.

The Physiome project, initiated in the 1990s with the vision of developing a comprehensive understanding of a healthy human~\cite{Hunter-2003-ID18918}, progressed further, and after 10 years it had already started incorporating genetic inheritance and environmental influences~\cite{Hunter-2010-ID19368,bassingthwaighte09}. In the first decade of this millennium, the Virtual Physiological Human (VPH) concept emerged~\cite{Hunter-2013-ID18920}. The Physiome project specifically, but also the wider domains of systems biology, personal health systems, biomedical informatics, and systems pharmacology, all face the common challenge of integrating complex and diverse data and models. Today, the focus has shifted from a singular model of ``the human'' to disease-specific models and variability within cohorts, akin to marking continuous transitions between the village and mountain views.
However, in certain aspects, the enthusiasm surrounding computational M\&S has proven to be overly optimistic. The importance of reliable and relatable high-quality data, as opposed to sheer data quantity, has become evident in the journey from the village to the mountain top. Considerations such as species-specific~\cite{Morotti-2021-ID18334}, sex-specific~\cite{Pierre-2022}, age-specific, and disease-specific modelling have been acknowledged, requiring more concerted efforts and coordination between in silico, in vitro, in vivo, and demographic data acquisition, processing, and sharing.

The advent of new experimental methods and the rise of machine learning~\cite{Siontis-2021-ID18107,Feeny-2020-ID14301,Attia-2021-ID17203,Nagarajan-2021-ID16577} have introduced an apparent dichotomy between mechanistic and data-driven approaches. Data-driven models seem to navigate a cloudy summit, while mechanistic models dwell in the village. Both possess the potential for integration, as data-driven models can be hypothesis proffering (mountain-to-village information transfer), while mechanistic models, especially when bridging multiple scales (temporal, spatial, species, disease\ldots) can offer relevant data input (village-to-mountain information transfer). With the latter being built upon first principles, they are more likely to generalise well and respect fundamental laws of physics. The former, however, are directly linked to real-world observations and thus more likely to capture important phenomena of in vivo (patho-)physiology. Looking to the future, we need to integrate data-driven and mechanistic modelling approaches, as visualised in Figure~\ref{fig:synergy}, more systematically in order to use the full potential of both.

The societal benefits of computational M\&S, outlined at the turn of the millennium~\cite{kohl00}, are increasingly tangible. From serving as teaching tools to aiding decision-making in clinical trials, computational models see adoption into practice. During the COVID-19 pandemic, M\&S (mostly data-driven) became invaluable~\cite{Sun-2023-ID18934}. When it comes to standardising and individualising medical care, there is still a long way to go, in spite of first success stories. The millennial promise of reducing morbidity and mortality turned out to be true: life expectancy in the EU rose from 80.9 in 2002 to 84.0 years for women in 2019 (before the pandemic) and from 74.3 to 78.5 years for men. However, we are not aware of data that would allow one to quantify the contribution of M\&S to this improvement. Accepted contributing factors include a reduction in infant mortality, rising living standards, improved lifestyles, and better education, as well as advances in healthcare and medicine in general~\cite{StatisticsExplained}.

The following sections provide perspectives on M\&S applications in basic and translational research, digital twins, in silico clinical trials, and machine learning.

\subsection{Utilisation of Modelling and Simulation in Basic, Mechanistic, and Translational Research}
The last few decades have witnessed improved cross-fertilisation between wet lab / clinical and computational methods in basic and translational research. Novel experiments and methods~\cite{Remme-2023-ID19242} have contributed to the generation of unprecedented data for M\&S applications in terms of both quantity and quality. In turn, in silico experiments have not only yielded novel, experimentally testable hypotheses but also served to falsify hypotheses derived from experimental or clinical observations~\cite{Gerach-NOGA}. The conceptual framework, mathematics, and technology underpinning and enabling models have evolved. Often, this evolution was iterative and continuous, for example enhancing resolution, expanding the "field of view" from tissue patches to whole organs, and extending time scales, but the field has also seen a number of disruptive changes, such as explicit representations of additional biological entities~\cite{EMI2021} and thermodynamically consistent formulations across multiple scales.
In cardiology, computer models have become widely employed~\cite{Niederer-2018-ID12269}, with atrial arrhythmias standing out as a particularly active area for research, translation, and clinical application, as reviewed recently.~\cite{Heijman-2021-ID18926}.

M\&S provides a highly controlled environment, facilitating the identification of cause-and-effect relationships. Confounding factors, often problematic to control or account for in wet lab settings (e.g., cross-talk between genetic and environmental effects), are more manageable in silico. Computational approaches demonstrate scalability by design, through parallelisation during execution of in silico experiments. Experiment design and setup as well as analysis of results represent the primary bottlenecks. An additional benefit of in silico research is the absence of inherent variability, obviating the need for repeated experiments. However, when explicit consideration of variability and uncertainty quantification are desired, multiple runs become imperative.
Computational approaches prove resource-efficient on multiple fronts by minimising human effort, reducing the burden on animals and the environment, as well as lowering financial and ethical costs.

\begin{figure}[!ht]
\centering\includegraphics[width=\columnwidth]{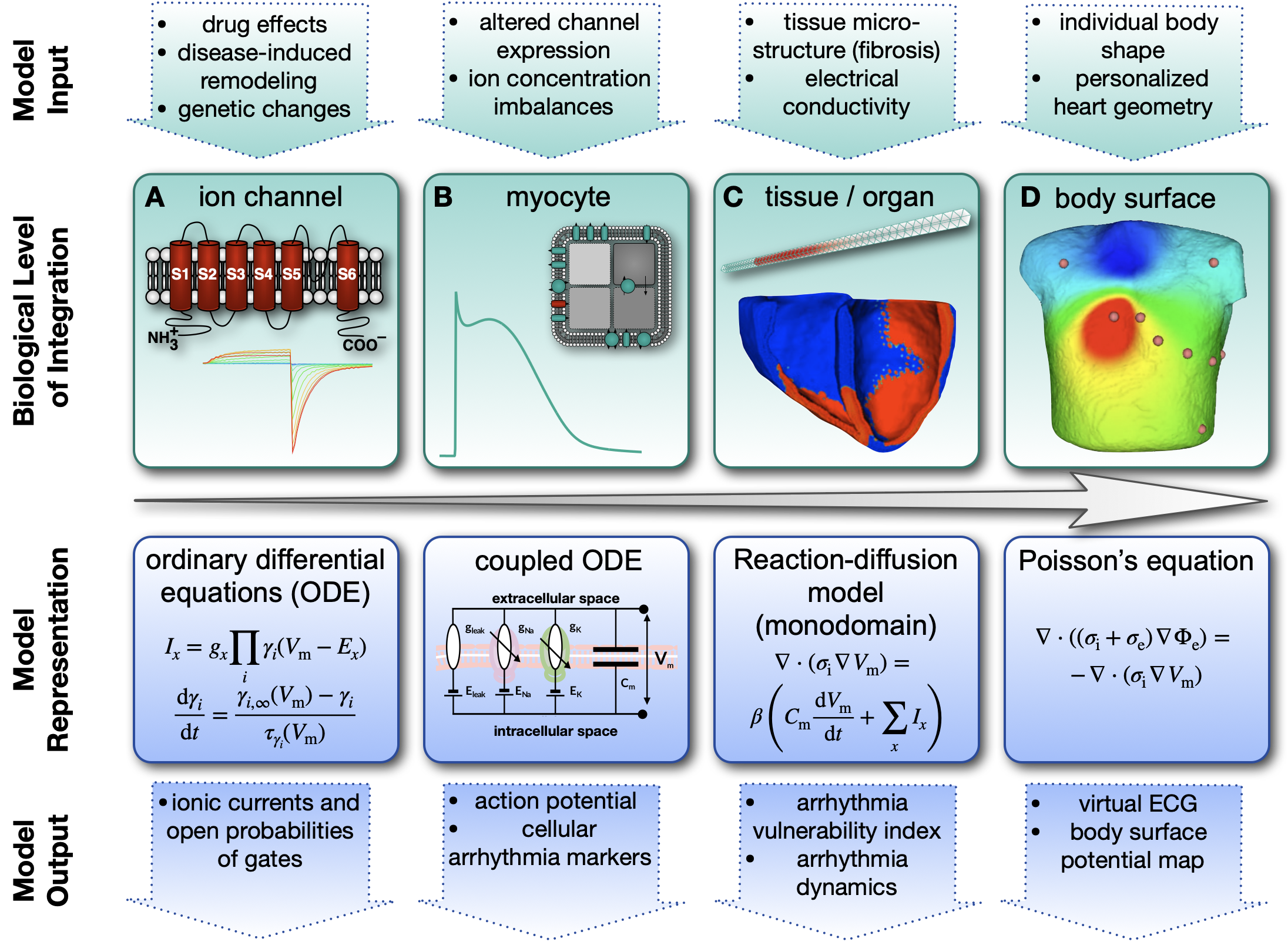}
\caption{Hierarchy of multiscale cardiac electrophysiology models ranging from ion channels (A) via
integrated cell (B) and tissue level models (C) to the body surface and electrocardiogram (D). The simulation system allows one to investigate what-if scenarios by changing input parameters of the model (top row) and analysing the effect on simulation outputs on numerous scales (bottom row) in comparison with wet lab and clinical data. Adapted with permission from~\cite{loewe18}.}
\label{fig:sim}
\end{figure}
As a concrete example, Figure~\ref{fig:sim} depicts a widely used multi-scale modelling framework in cardiac electrophysiology. Beginning at the smallest scale with single ion channels, their kinetics are described using ordinary differential equations (ODE), accounting for factors like ion concentrations and transmembrane voltage ($V_\mathrm{m}$) (Figure~\ref{fig:sim}A). These ion current models can incorporate effects of genetic mutations, drug effects, or altered experimental conditions. Moving to the cell level (Figure~\ref{fig:sim}B), electrophysiological models consider the various ion channels present in cardiac cell membranes. Represented by a system of coupled ODE, these models yield action potentials and can be adjusted to reflect different cell states, for example during disease-induced remodelling. In simulation studies, myocyte models are frequently emphasised. However, it is essential to recognise that the heart contains more non-myocytes (such as fibroblasts and macrophages) than cardiomyocytes. Models also exist for most of these non-myocyte cell types~\cite{Xie2009,SimonChica2021,Seemann-2017-ID11878}.
As excitation propagates through cardiac tissue, spatio-temporal changes in $V_\mathrm{m}$ occur (Figure~\ref{fig:sim}C). This coupling can be mathematically represented using reaction-diffusion systems with partial differential equations (PDE). This approach allows one to simulate activation wavefront propagation and to integrate factors like anatomical variability or fibrosis in personalised models.
Local differences in $V_\mathrm{m}$ generate currents that create an electric field, described by Poisson's equation. This field extends to the body surface, enabling the acquisition of virtual electrocardiograms (ECG) (Figure~\ref{fig:sim}D).
Throughout these scales, model parameters can be adapted for example to represent diseases, to create digital twins for individual patients or digital chimeras representing a likely virtual patient sample from a specific population, or to simulate therapeutic interventions (Figure~\ref{fig:sim} top row). Model outputs can be evaluated on all represented biological integration levels (Figure~\ref{fig:sim} bottom row), and also be compared to wet lab and clinical data.

\subsection{Digital Twin Approaches}
\begin{figure}[!ht]
\centering\includegraphics[width=0.8\columnwidth]{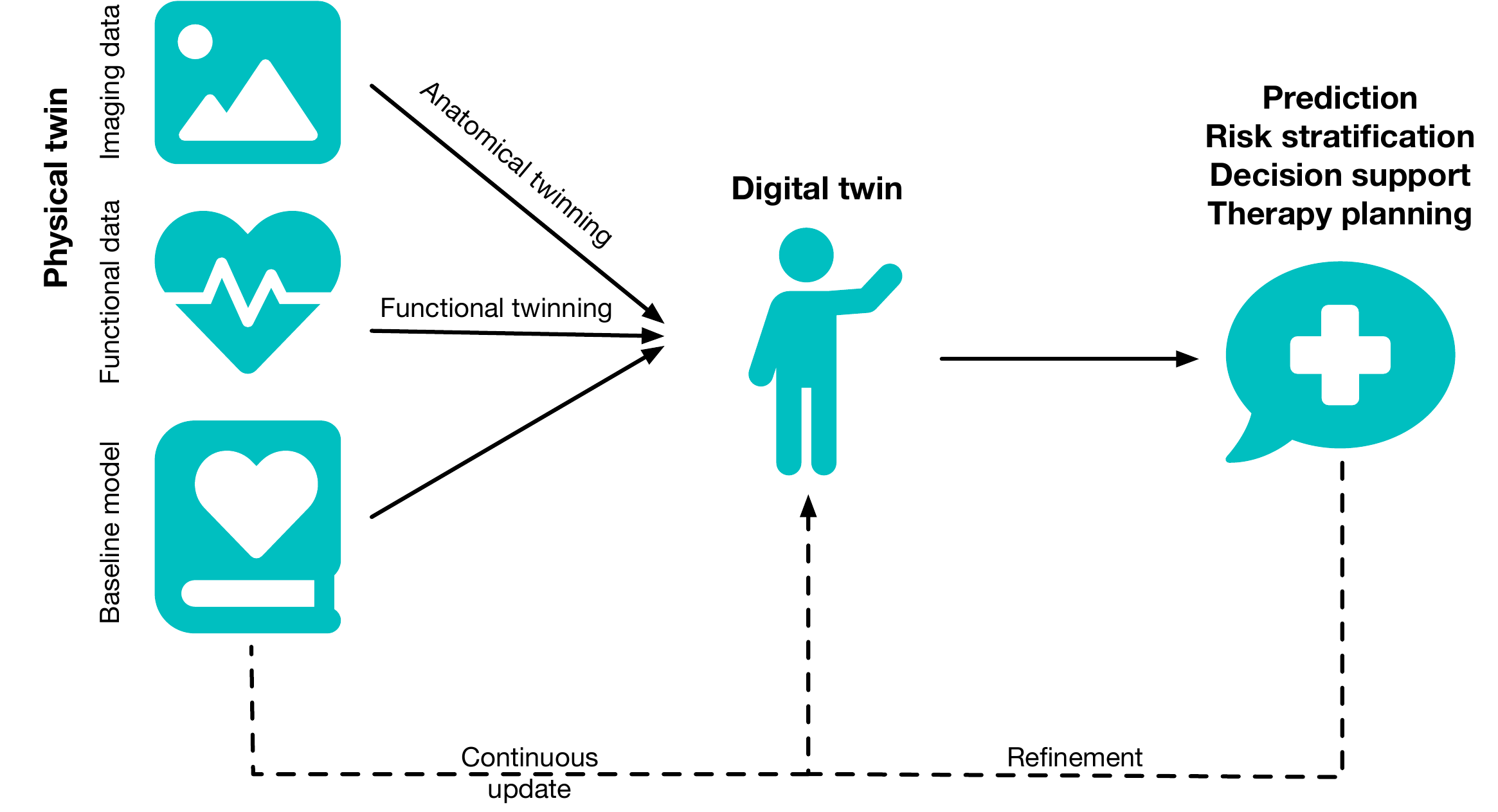}
\caption{Digital twin workflow. A baseline model builds the basis for the digital twin. Often, it is a bottom-up mechanistic model, informed by biophysical first principles and population-level knowledge. Anatomical and functional personalization are performed based on individual patient measurements (e.g., computed tomography or magnetic resonance imaging for anatomical twinning and ECG for functional twinning). The parameterised digital twin can then be used in computational simulations to make predictions regarding personal risk or support decisions regarding optimal therapy for personalised medicine. The digital twin should be updated continuously
when new measurements are available and refined by comparing predictions to real world outcomes. Adapted with permission from~\cite{Jadczyk2021a}.}
\label{fig:twin}
\end{figure}

\begin{figure}[!ht]
\centering\includegraphics[width=\columnwidth]{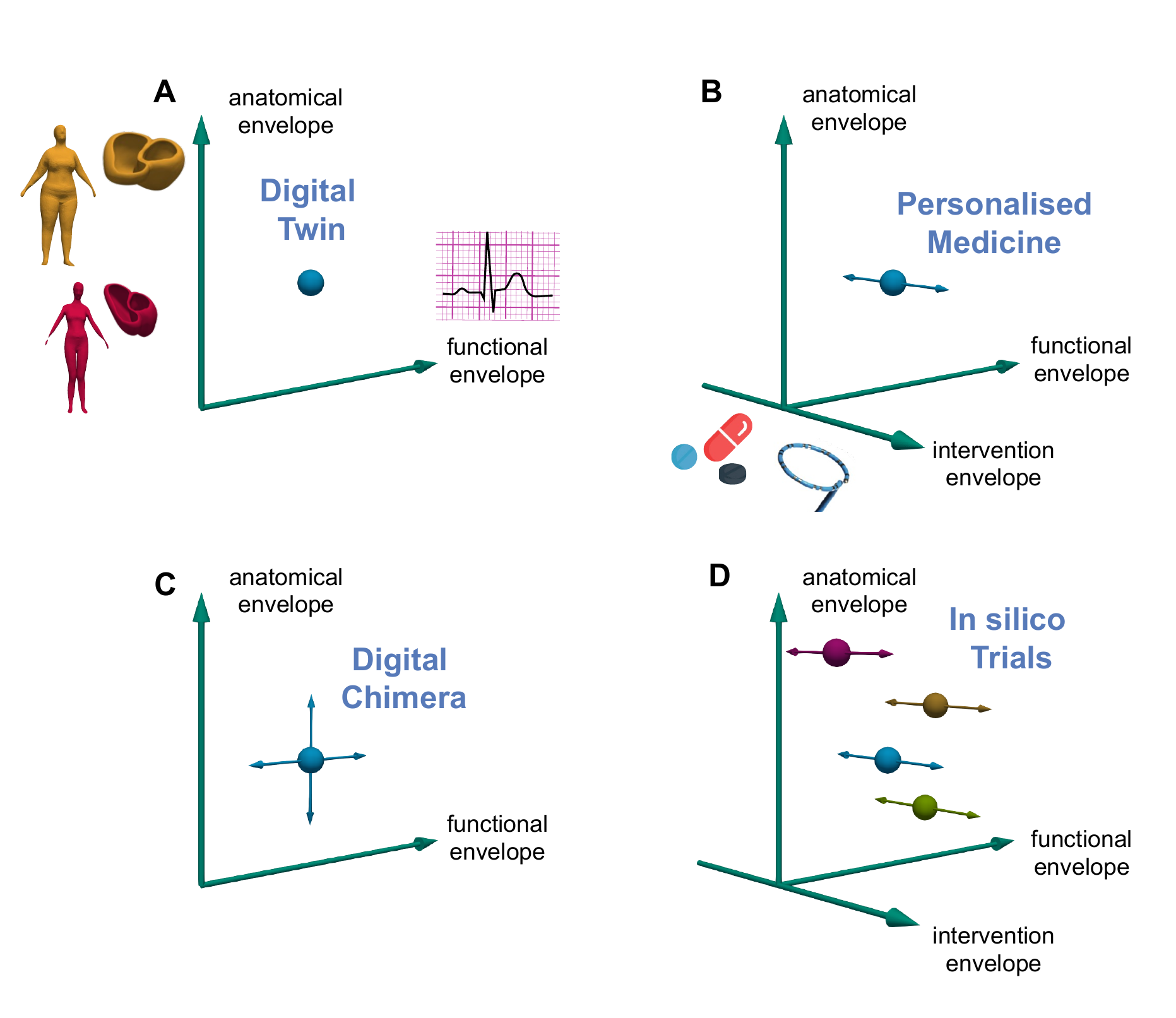}
\caption{Computational M\&S approaches. Representing an individual as well as possible with a computational digital twin model requires both anatomical and functional personalization (A). Once a digital twin is established, different interventions can be evaluated in silico, for example to support precision medicine (B). If anatomical and functional envelopes are continuous and limited to biologically relevant ranges, arbitrary numbers of new virtual subjects (digital chimeras) can be sampled from this space, representative of a population but not of a specific individual in this population (C). With a whole cohort of digital twins, digital chimeras, or a mix thereof, in silico clinical trials can be conducted, evaluating the effects of specific interventions consistently applied to the entire virtual study population (D). Figure inspired by Zeike A. Taylor, University of Leeds.}
\label{fig:concepts}
\end{figure}

A digital twin is a personalised computational model of an individual patient, mimicking various aspects of their structure and function. More generally, this approach can also be a digital representation of a technical object (e.g., a cable car shuttling between the village and the mountain top or certain components of it)~\cite{aiaa2020digital}. Though often applied for decision support in critical time steps, digital twins, in general, are dynamically updated throughout their life-cycle from the physical twin (Figure~\ref{fig:twin}). If such continuous, bidirectional exchange of information between the physical and the digital twin is not implemented, the terms ``digital shadow'' or ``digital snapshot'' have been proposed.
Currently often focused on a specific organ, the geometric representation of a personalised computational model is derived from imaging data of the patient's unique anatomy, which in itself often encodes diagnostic information~\cite{Govil2023,Jia2021}. By adjusting and optimising functional model parameters based on measured clinical data, the model aims to capture the physiology of the patient's organ or of multiple organs in silico. While more and more comprehensive models and personalisation strategies become available, models will and should always be a simplification of reality. The law of parsimony, also known as Occam's razor, recommends searching for explanations constructed with the smallest possible set of elements. Figure~\ref{fig:concepts}A illustrates these aspects of anatomical and functional twinning.

Digital twins are used for individual risk prediction, decision support, and therapy planning~\cite{Jadczyk2021a,Corral-Acero-2020-ID13459,Niederer-2021-ID16335}. For example, different options for interventions can be evaluated in silico before deciding on the approach to be applied to a patient, in keeping with the personalised medicine vision (Figure~\ref{fig:concepts}B). While digital twins have proven valuable across diverse medical applications, challenges remain, particularly in automated and robust generation of personalised computer models using clinical data. Addressing these will be crucial for the application of digital twins in broader clinical practice.

\subsection{In Silico Clinical Trials}
In silico clinical trials use computational M\&S for evaluating safety and/or efficacy of a medical procedure or product, including drugs and devices~\cite{Viceconti-2016-ID18928}. In silico clinical trials are a novel and potentially disruptive methodology that can be applied to a multitude of Contexts of Use including reduction, refinement, and replacement of in vitro, animal, and human experiments~\cite{Viceconti-2021-ID15999}. Similar to laboratory experiments and conventional clinical trials, the specific Context of Use, i.e., ``the specific role and scope of the computational model used to address the question of interest''~\cite{v2018assessing}, needs to be clearly defined a priori~\cite{GSP}. Encouragingly, evidence generated through M\&S can be used during the regulatory qualification process~\cite{FDA,frangi_2023_8325274}. 

In silico clinical trials can be based on a sizeable cohort of digital twin models~\cite{Sarrami-Foroushani-2021-ID18927} but creating them at scale is non-trivial, both in terms of effort and data availability. An attractive alternative is the use of methods that allow one to create ``new'' virtual individuals --- digital chimeras (Figure~\ref{fig:concepts}C). These are not a digital replica of a specific patient, but rather represent properties of a population well in a statistical sense. For the anatomical dimension, statistical shape models are frequently used to draw samples from a continuous space that matches the statistical distribution of the samples that were used to build the statistical shape model~\cite{Ambellan-2019-ID19245,Dou2024}. For the functional dimension, populations of models can be built from bounded, high-dimensional parameter spaces, for example with Latin hypercube sampling~\cite{britton13}, aiming to match empirically observed distributions of simulation output characteristics~\cite{MedalCare-XL,Lawson-2018-Unlockingdatasets}.

Once the in silico study cohort is established, each virtual participant (either chimera or twin) is subjected to a number of predefined interventions (Figure~\ref{fig:concepts}D). In contrast to biological participants, multiple interventions can be tested in the same in silico individual, starting from the exact same baseline conditions, potentially yielding more meaningful control and reference data than would be possible in wet lab and clinical settings. 

\subsection{Machine Learning Enabled by M\&S}
Machine learning, a subset of artificial intelligence, emulates human learning processes by leveraging data, algorithms, and statistical models to make predictions without explicit instructions. In medicine, machine learning algorithms now rival or even surpass human capabilities in certain tasks, especially in fields like medical imaging~\cite{Sermesant-2021-ID16038} and cardiac electrophysiology~\cite{Jadczyk2021b,Trayanova-2021-ID15833}.

Despite the apparent abundance of data, machine learning for biomedicine faces a number of challenges including the "big data but small data" paradox. Root causes for this issue are that, in practice, data are not readily available for legal, ethical, or technical reasons (or misconceptions~\cite{Gefenas2021}), and that data are not sufficiently standardised or curated~\cite{DataDGK}. Ground truth datasets with ample size and high-quality labels are frequently absent. Additionally, biased and unbalanced datasets pose questions about capturing relevant data ranges densely enough and about ensuring fairness in the developed machine learning algorithms~\cite{Puyol-Anton-2021-ID17011,pilia20,Jones-2024-ID19369}.

To enhance machine learning applications in biomedical settings, several potential solutions are being explored:
unsupervised or semi-supervised learning to improve data annotation, data augmentation techniques to expand datasets, automation of dataset linkage (learning relations rather than manually defining them), federated learning across multiple institutions without the need of data sharing~\cite{Rieke-2020-ID18106}, encouraging and facilitating data donation, and incentivising publication of FAIR (findable, accessible, interoperable, reusable) and ideally open data, metadata, and code~\cite{Wilkinson-2016-ID13362,FAIR4RS}.
Computational M\&S is one of the fields with vast potential to tackle the above-mentioned problems by integrating machine learning and multiscale modelling to leverage synergies between the two approaches~\cite{Alber-2019-ID18913,McDuff-2023-ID18551,Corral-Acero-2020-ID13459,Lyon2019}. Combination of machine learning and M\&S methods can allow for better phenotyping and stratification of diseases~\cite{Lyon2019}. Furthermore, many of the legal and ethical issues do not apply to synthetic data, and ground-truth annotations can be assigned with very high certainty in most M\&S approaches. Using appropriate sampling schemes, the balance of classes (e.g., healthy controls and different disease subtypes) and parameters (e.g., age, sex, obesity) can be controlled. In multi-scale models, data augmentation can be lifted to higher levels by enabling mix \& match, for example combining $N$ heart models with $M$ torso models, resulting in $N \times M$ permutations. In such approaches, co-variances should be accounted for~\cite{Dou2024}. In silico generated data can serve as training data (either purely synthetic~\cite{giorgioML,Pilia-2022-ID18150} or hybrid~\cite{giorgioEP,Sanchez-2021-ID16474} when combined with real-world recordings) for developing machine learning algorithms that are then evaluated on real-world signals. In keeping with this, many open benchmark datasets with well-defined ground truth are synthetic.

In spite of its huge potential, the use of M\&S for generating machine learning training data also introduces specific challenges. One needs to ensure model fidelity (i.e., accurately reflecting real-world scenarios) through rigorous and use case-specific validation. Difficulties include the identification of an upper limit of predictive error for any relevant real-world input data beyond the calculation of the predictive accuracy over a finite number of observed values, as well as the ``plausibility trap'': model outputs (mechanistic or data-driven) that match observed biological behaviour are not necessarily a proof for having found (explicitly or implicitly) the underlying causal link~\cite{Quinn-2011-ID19517}. Potential domain gaps between simulated and real-world data can also be due to model oversimplifications. As an example: if the model does not include the signal acquisition process, the simulated data will not contain measurement noise. To be suitable for training a machine learning algorithm that is hoped to perform well on noisy real-world data, these synthetic data will need to be artificially 'corrupted' with realistic noise~\cite{PVC}. Another aspect is that simulated data are not guaranteed to cover the relevant variability by default. As an example, one needs to consider the diversity in the dataset used to generate a statistical shape model when aiming to represent a diverse population well. Any machine learning approach will only be as good as the training data, which causes problems when uncommon cases are under-represented. In addition to cases not adequately represented in the training set, there can also be ``silent variables'' that do not significantly affect the specific quantity of interest and thus limit the generalisation capacity of the model. For example, the variable age might not be needed to predict myocardial infarction risk in a given dataset used to develop the machine learning model. However, for more general out-of-sample application, the predictor might perform poorly when not taking age into account.

Despite examples of remarkable classification power of purely data-driven approaches, we believe that knowledge adds value as a way to understand the true nature of the problem rather than being left with a fragmented landscape of bits of insight. While the field of explainable artificial intelligence (xAI) tries to address this shortcoming, current approaches only provide insight into which input data are essential for the decision, rather than why they are (in many cases). Machine learning that is constrained by the laws of physics or mechanistic models of biological function can be a way of synergistically combining data-driven and mechanistic approaches. 

One promising method to achieve this combination is the use of physics-informed neural networks that offer advantages regarding consistency with physics, data efficiency, extrapolation capacity, and applicability to inverse problems. This emerging class of tools has already been used for a wide range of applications in cardiovascular M\&S~\cite{RuizHerrera-2022-ID18046,Caforio-2023-ID18896,Martin-2021-ID17251,SahliCostabal-2020-ID13484}. Another way of merging mechanistic M\&S with data-driven approaches is the use of statistical emulators that reduce computational effort when calculating specific quantities of interest and thus allow for parameter identification combined with uncertainty quantification and sensitivity analysis including higher-order interdependencies~\cite{Coveney-2018-ID12271,Rodero-2021-ID16098,Strocchi-2023-ID18664,Regazzoni-2021-ID16493}.

\section{Tools \& Ecosystems}
The monumental task of constructing the human Physiome can only be achieved by a collaborative effort involving the integration of numerous modules in open and interoperable ecosystems. These modules comprise for example single-organ and single-physics models, data repositories, as well as pre- and post-processing pipelines. Key prerequisites for enabling them to work together include the establishing of standards that aid interoperability and the sharing of data, metadata, and code. 
As these principles align naturally with the notion of Open Science, its rise is no surprise. 
This rise is driven by the recognition that collaboration, transparency, and shared resources are essential for the progress of modern day research and development. Open Science fosters reproducibility, increases efficiency, and reduces redundant efforts.

Physiologically detailed computational M\&S heavily relies on experimental data for the design, parameterization, and validation of models. Adherence to the FAIR principles is crucial for getting closer to the realisation of the Physiome vision and ideally data and metadata should both also be open~\cite{DataDGK}.
To facilitate implementation, ``Ten Simple Rules for FAIR Sharing of Experimental and Clinical Data with the Modeling Community'' were formulated~\cite{KoenigFAIR}. Additionally, the modelling community often gives back results of anatomical model building~\cite{atrialshape,Strocchi-2020-ID14465} and simulation studies~\cite{MedalCare-XL} and has demonstrated the benefits associated with wide use, especially when using standardised reference frames~\cite{Schuler-2021-ID16646,Roney-2019-ID14879,Nairn-2023-ID18842}. Subsequent studies were enabled through sharing these models, the original author got credit by citing the data publication, and society benefits from the new insights that might otherwise not have been obtained at all or at a higher cost.
Despite commendable efforts to reform research assessment, like DORA~\cite{DORA} and CoARA~\cite{COARA}, the lack of proper incentive and recognition systems for sharing models, data, metadata, and code remains a substantial hindrance to implementation of open science in general, and the FAIR principles in particular..

Besides data and metadata, software is crucial for M\&S. The FAIR principles for research software (FAIR4RS) have only just been formulated (in 2022~\cite{FAIR4RS}), even though the essential role of software has long been acknowledged~\cite{Anzt-2020-ID13991}. Thriving M\&S ecosystems can emerge based on FAIR research software~\cite{openCARP,Africa-2023-ID18725,Cooper-2020-ID17200}, workflows~\cite{Williams-2021-ID18914,cobiveco,peerp}, and collaborative development environments~\cite{cde-fdm, Houillon_Klar_Stary_Loewe_2023}.

For data and code, standards are an essential requirement, enabling interoperability and modularisation~\cite{Schreiber2020,Brunak2020}. Examples include CellML~\cite{Clerx-2020-ID14875}, SBML~\cite{Hucka-2003-ID18919}, FieldML~\cite{Christie-2009-ID18933} and the SPARC Data Structure (SDS)~\cite{Bandrowski2021}, as well as numerous domain-specific markup languages catering for specific research domains~\cite{trevisanjost:EP23,Gsell2023}. Often building on community conventions, international standards were established including ISO TS 9491:2023 (Recommendations and requirements for predictive computational models in personalised medicine research), ISO 20691:2022 (Requirements for data formatting and description in the life sciences), ISO 23494 (Provenance information model for biological material and data), and ASME V\&V40 (Assessing credibility of computational modeling through verification and validation: application to medical devices).
A common language and clear interfaces can then enable large collaborative efforts like the 12 LABOURS Digital Twin Platform~\cite{BabarendaGamage-2023-ID18911}, or decentralised portfolio approaches as envisioned in the European Virtual Human Twin Manifesto~\cite{VHT} and roadmap~\cite{VHTRoadmap}. 

Besides technical harmonisation, community building has been a driver for sustained impact in the last decades. The Virtual Physiological Human community, coordinated by the VPH Institute~\cite{Hunter-2013-ID18920}, the Computational Modeling in Biology Network (COMBINE)~\cite{Schreiber2020}, or software-centred communities~\cite{Updegrove-2017-ID19252,openCARP,Cooper2020} are just a few examples. The International Union of Physiological Sciences (IUPS) took a leading role in this field by driving the Physiome project, building and operating the Physiome model repository~\cite{Yu2011}, organising symposia, and co-publishing the Physiome journal.  

In conclusion, tools and people are essential for thriving and sustainable in silico ecosystems. While tool and data sharing can be tackled by the community through technical means, ill-developed recognition mechanisms and the frequent disconnect between project funding cycles, which are limited by definition on the one side, and ambitions and expectations for sustained software development and support impact on the other side form challenges to which new solutions must be found by academia, industry, regulators, and politics. Both systems and communities need support to remain functional and valuable, as already identified in the 1970s in Lehman's laws of software evolution~\cite{Lehman1980}. Educational structures are also rarely aligned well enough with the needs of interdisciplinary research in general, and the emergence of disruptive technologies (such as AI-based M\&S) in particular. When training the next generations of scientists, too few programmes aim at developing a truly diverse set of skills covering wet lab, in silico, and clinical content~\cite{DGKwhite}.

\section{Cardiology Applications}
Cardiology stands out as a field with particularly active and successful M\&S activities, as reviewed in detail elsewhere~\cite{Niederer-2018-ID12269,Trayanova-2023-ID18909,quarteroni17}. Here we focus on select 21\textsuperscript{st} century examples to provide the interested reader with starting points for further exploration of cardiac M\&S.
Through these spotlights, we aim to showcase the versatility and impact of computational modelling in addressing complex problems across different facets of cardiology. The examples not only underscore achievements of M\&S in this field, but hopefully also serve as inspirations for further exploration and innovation within the broader landscape of cardiovascular research.

The cardiac ``function'' with the longest history of computational M\&S is electrophysiology. Soon after the establishment of the seminal Hodgkin-Huxley model for the electrophysiology of neurons, Noble adapted it to Purkinje cells in the heart (in 1962~\cite{noble62a}). 
Presently, models implementing the concepts introduced in Figure~\ref{fig:sim} are largely rooted in those early works and build on their principles.  To strike a balance between physiological detail and computational cost, models can be deliberately reduced or embedded in model hierarchies with increasing complexity~\cite{NagelProp}. Cardiac electrophysiology models find routine application in arrhythmia~\cite{Trayanova-2023-ID18909} and ischaemia research~\cite{loewe18,loewe15}, for ECG simulations~\cite{ECGreview}, and they hold promise for personalising ablation therapy~\cite{Boyle-2017-ID11783} as well as for accelerating drug development~\cite{Musuamba-2021-ID16400}. 

Ultimately, the heart's main function is that of a mechanical pump, which has also been replicated in silico~\cite{Chabiniok-2016-ID18912} including material properties~\cite{Emig-2021-ID18343} and growth~\cite{Goktepe-2010-ID13429}. Electrical and mechanical functions are coupled bidirectionally through excitation contraction coupling and mechano-electric feedback~\cite{Quinn-2021-ID14878}. This interplay has been analysed in detail in several simulation studies using electro-mechanically coupled models~\cite{Gerach-2021,Fedele-2023-ID18538,Augustin-2016-ID11752,Gerach-2024-ID18940}. Blood flow through the heart and the circulatory system can be studied in silico using computational fluid dynamics and fluid structure interaction models~\cite{Bucelli-2023-ID18202,brenneisenCoupling21,Schwarz-2023-ID18915,Zingaro-2022-ID18916}.

Since the turn of the millennium, M\&S has been used for numerous cardiological Contexts of Use~\cite{Viceconti-2021-ID15999}. In mechanistic and basic research, simulation studies helped for example to better understand electrogram~\cite{Sanchez-2022-908069} and ECG genesis~\cite{moss21,loewe16e,Andlauer-2019-ID12169}, disease mechanisms~\cite{Loewe-2019-ID12801,Antoine} as well as pharmacological modes of action~\cite{loewe14b,Wiedmann-2020-ID14070,Passini2017}. The CiPA initiative was launched in 2013 to engineer an assay for assessment of the proarrhythmic potential of new drugs with improved specificity compared with the hERG assay plus Thorough QT study~\cite{Sager-2014-ID19367}. This example --- steered by international regulators, industry, academics and nonprofit organisations --- shows how support by key stakeholders including regulators can accelerate and enhance impact~\cite{Strauss2021}. The latest version of the ICH Guideline ``Clinical and Nonclinical Evaluation of QT/QTc Interval Prolongation and Proarrhythmic Potential'' (ICH E14/S7B) explicitly encourages the use of in silico models to integrate experimental ion channel data for clinical and preclinical research and development.

A wide range of digital twin approaches has been proposed for personalised ablation therapy~\cite{Azzolin-2023-ID17490,Solis-Lemus-2023-ID18614,Gerach-2021,Gillette-2021-ID16142,Loewe-2019-ID12386,Niederer-2020-ID14134,Roney-2023-ID18891,Das2023} and the first prospective clinical trials involving modelling-derived predictions report favourable results~\cite{Shim-2017-ID11782,Boyle-2019-ID12859}.

Cardiac in silico clinical trials have been reviewed elsewhere~\cite{Rodero-2023-ID18638} and include virtual ablation studies~\cite{Gerach-Ablation,PersonAL,Gharaviri-2021-ID18923}, culminating in a holistic benchmark setting for mapping-guided ablation, covering the full process from catheter positioning and deformation via signal acquisition, filtering, and processing to the selection of ablation sites~\cite{Alessandrini-2018-ID12168}. The well-controlled in silico setting has been used as a complementary approach to integrate and consolidate clinical trials~\cite{Lehrmann-2018-ID11871} and for systematic evaluation of newly developed or refined medical devices~\cite{Monaci-2022-ID18328} and algorithms~\cite{Frisch-2020-ID13714,vilaAbl}.
In the field of cardiac pharmacology, in silico methods support the evaluation of drug efficacy and safety in sizeable cohorts of several hundred virtual patients, often following the digital chimera approach described in Figure~\ref{fig:concepts}C~\cite{Dasi-2023-ID18696,Trovato-2022-ID17522,Peirlinck-2022-ID17588}.
Similar approaches have also been used to study hemodynamics~\cite{Mill-2021-ID16427}.

Several of the machine learning breakthroughs seen in recent years in the cardiology field~\cite{Trayanova-2021-ID15833} were enabled by M\&S, which highlights the potential for synergy between data-driven and mechanistic models. The possibility to augment datasets on specific anatomical and functional levels (Figure~\ref{fig:concepts}) was used, for example, for training classifiers to predict the acute success of pulmonary vein isolation as a treatment of atrial fibrillation~\cite{giorgioML}, for distinguishing different types of atrial flutter~\cite{giorgioEP}, and for localising the origin of ventricular ectopic beats~\cite{PVC,Doste-2022-ID18105}. All these examples relied on multi-scale simulation of cardiac electrophysiology ranging up to the ECG as provided in the open MedalCare-XL dataset~\cite{MedalCare-XL}, comprising 16,900 healthy and pathological synthetic 12 lead ECG recordings. Better control of the distribution of samples across classes (e.g., balanced size of control and disease cohorts) helped to improve the diagnosis of fibrotic atrial myopathy~\cite{NagelFAM}, left atrial enlargement~\cite{Nagel-2022-ID17346}, and electrolyte imbalance~\cite{pilia20}. In some cases, the synthetic training data were augmented with a small number of clinical recordings, resulting in a hybrid training data set~\cite{giorgioEP,SanchezML,Roney-2022-ID17385,Shade-2020-ID14203}.

\section{Outlook}
As M\&S continues to develop at the intersection of computational advances and biomedical exploration, the outlook for the future is exciting. The rapid evolution of M\&S techniques we have witnessed since the turn of the millennium, coupled with the continuing progress in machine learning and Open Science practices, has started to propel the field to new frontiers. In the forthcoming years, the application of computational approaches in cardiology and beyond is poised to deepen, addressing increasingly complex questions with high precision. M\&S can serve as a catalyst for the advancement of digitally controlled wet labs and contribute to tightly linked and fast iterations between wet lab experiments, big data analysis, conceptual hypotheses, mechanistic models, in silico plausibility testing, and comprehensive evaluation, leading to new hypotheses that can be experimentally validated and to refined experimental designs partially exploiting automated AI-driven laboratories for compound synthesis. This has heralded a revolution in drug design and testing reducing time and cost~\cite{Zhavoronkov2019}. Robust standards, data \& code sharing, curation, and incentivisation are required to stay on a dynamic trajectory of M\&S. This Outlook section highlights essential challenges in three application areas that need to be addressed to yield innovations that will shape the future landscape of M\&S in biomedical research.

\subsection{Frontiers for Computational Models}
Concerning model formulations, current challenges include the requirement to improve multi-physics models. Existing electromechanical models~\cite{Gerach-2021,Augustin-2016-ID11752} and fluid mechanical models~\cite{Bucelli-2023-ID18202} are on the way of being extended to include perfusion~\cite{Zingaro-2023-ID18528}, growth~\cite{Tikenogullari-2023-ID18924}, and metabolism. Maintaining biophysical and energetic consistency not only across scales but also across ``functions'' is not trivial. This becomes relevant as well when extending models towards even more microscopic scales like the extracellular-membrane-intracellular model for electrophysiology~\cite{Tveito-2021-ID18922,RosilhodeSouza-2024-ID18921} potentially also including multiple cell types and extracellular structures. Another continuous issue is model validation for additional Contexts of Use~\cite{Pathmanathan-2018-ID18932,Aldieri-2023-ID18929,johnstone16}, often complicated by scarce experimental data or limited access, also concerning control data from internal organs in human.

The breadth of methods applied in the field of M\&S poses ever-growing requirements on the training of researchers. The body of relevant knowledge continues to grow in all disciplines involved and skills ranging from finite element analysis via machine learning to physiology and medicine are essential to use the full potential of in silico methodologies. Stringent time limits in many phases of training, ranging from undergraduate studies to postdocs, can amplify the problem. As a single individual can hardly be trained in all relevant fields, improved communication across disciplines may be the only viable and attractive solution. It will be interesting to see whether larger labs or intensified collaboration will the better way, but we need to learn and practice how to communicate across disciplines: translation needs translators!

The advent of generative artificial intelligence and foundation models including large language models offers unprecedented potential for even tighter integration of data-driven and mechanistic models and may further blur the boundaries between them. To capitalise on this potential, tools and interfaces for bidirectional data exchange between those village and mountain-top views will be required, combined with novel concepts for validation of model predictions.

\subsection{Frontiers for Advancing Digital Twins}
When using M\&S to facilitate precision medicine through digital twins, personalisation of generic baseline models is a major task. Estimating model parameters from clinical and/or experimental measurements often involves solving inverse problems. This can usually only be achieved with limited temporal and spatial resolution and remaining uncertainty~\cite{unger21erp,kovacheva20,Emig-2021-ID18343,Coveney-2022-ID18047,coveney,Camps-2023-ID18665}. When integrating and assimilating multiple data sources, models can become over-constrained so that data and model fidelity need to be balanced~\cite{Pagani-2021-ID15852}, a process in which machine learning approaches like physics-informed neural networks can be valuable tools~\cite{Caforio-2023-ID18896}. How input parameter uncertainty and variability influence simulation outputs after being propagated through a model calls for dedicated analyses~\cite{mirams16,Gray-2018-ID18931,Clayton-2020-ID14136,Winkler,Strocchi-2023-ID18664,Karabelas-2021-ID17109,Bergquist-2023-ID18804}, for example using  uncertainty quantification tools~\cite{Tate-2023-ID18917}. 

While the concept of a systematic digital representation of human pathophysiology, a comprehensive digital twin, has been under consideration for many years, existing research primarily focuses on specialised patient-specific models predicting specific clinical entities, often at single points in time and/or limited to single organs~\cite{Viceconti2024}. The development of comprehensive digital twins faces multifaceted challenges including scientific, technical, ethical, legal, and cultural aspects, as outlined in the Virtual Human Twin Manifesto~\cite{VHT}. A recent draft for a Virtual Human Twin roadmap~\cite{VHTRoadmap} details specific actions, including the development of a data repository, intellectual property management, incentives, regulatory clarity, clinical evidence generation, and universal access to digital twin technology in healthcare. Despite the efforts for more and more comprehensive models, we should not forget the value of the simplification and abstraction that models offer (by definition), which can be advantageous for enabling mechanistic insight in knowledge-generating research.

\subsection{Frontiers for Advancing in silico Clinical Trials}
In silico clinical trials rely on big cohorts, such as those included in the UK Biobank~\cite{Bycroft2018}. In practice, scaling up digital twin model generation (Figure~\ref{fig:concepts}A) to this range remains challenging~\cite{Niederer-2021-ID16335}. Thus, methods to digitally sample specific aspects of the models from underlying biological distributions are attractive to better control variability using a limited set of real world samples. While digital chimeras (Figure~\ref{fig:concepts}C) as the individuals in synthetically derived populations of models / in silico cohorts (Figure~\ref{fig:concepts}D) can be informed by parameter ranges~\cite{britton13} or statistical shape models~\cite{atrialshape,rodero_2021_4506930}, identifying the underlying probability distribution for a target population without bias remains non-trivial~\cite{Puyol-Anton-2021-ID17011,LaMattina-2023-ID18930}. This is particularly challenging for high-dimensional parameter spaces with non-obvious interrelations of parameters, which need to be captured to constitute physiologically relevant virtual subjects~\cite{Dou2024}. If balanced data are not readily available, such as can be the case for (often under-reported) sex-specific or ethnicity-related differences~\cite{Pierre-2022}, computational M\&S can help to mitigate imbalances~\cite{Hellgren-2023-ID18925}.

\section{Conclusion}
The evolution of computational modelling of biological systems since the turn of the millennium has been nothing short of remarkable, ushering in mature applications across basic and clinical research. Equally remarkable is how much the visions for societal benefit, presented at the turn of the millennium~\cite{kohl00}, still hold today. While predictions in some fields might seem pragmatically optimistic when looking back, we also started seeing results of M\&S adding value in many areas of basic and applied research. After the historical debate on ``reductionism \& integrationism'' has all but subdued~\cite{Kohl2009}, a fresh dichotomy between mechanistic and data-driven modelling takes centre stage. Regardless of whether one sees the associated developments as challenges or opportunities, the dynamics of recent progress suggest that we may increasingly be able to combine the mountain and village views of research. Shifting focus from universal models to individualised representations including digital twins and digital chimeras for in silico basic research and clinical trials opens up exciting new avenues for future exploitation. Tackling the challenge of disease-specific models that take into account demographic information, intra- and inter-individual variability, uncertainty quantification, and data standardisation necessitate open collaboration, concerted community efforts, and the use of reliable tools and standards. The tightly controlled environment offered by in silico methods has become integrated into physiological and medical research, including multi-physics modelling, multi-scale studies, virtual cohort simulations, and machine learning. The synergy between mechanistic and data-driven modelling may become a powerful force to drive the next wave of artificial intelligence applications. This holds the promise of reshaping our understanding of physiology and disease mechanisms, fostering the generation and evaluation of innovative hypotheses, and ultimately contributing to the ongoing progress of preventive, diagnostic, and therapeutic approaches.

\paragraph{\textbf{Acknowledgments:}} This work was supported by the Leibniz ScienceCampus ``Digital Transformation of Research'' with funds from the programme ``Strategic Networking in the Leibniz Association'', the European High-Performance Computing Joint Undertaking EuroHPC under grant agreement No 955495 (MICROCARD) co-funded by the Horizon 2020 programme of the European Union (EU) and the German Federal Ministry of Education and Research (BMBF), as well as the German Research Foundation Collaborative Research Centre SFB1425 (\#422681845) and the NZ Government ``12 Labours'' MBIE grant. The authors acknowledge helpful feedback on the manuscript by Carmen Martínez Antón, Joshua Steyer, and Patricia Martínez Díaz.

\paragraph{\textbf{Conflicts of interest:}} All authors declare that they have no conflicts of interest.

\paragraph{\textbf{Author contributions:}} Conceptualisation: all authors; Writing – original draft: AL; Writing – review \& editing: all authors

\paragraph{\textbf{Acknowledgments:}} During the preparation of this work, the authors used ChatGPT for drafting and improving linguistic quality of the text. After using this service, the authors reviewed and edited the content as needed and take full responsibility for the content of the publication.


\bibliographystyle{RS}
\bibliography{references.bib}
\end{document}